\documentclass[twoside]{article}
\usepackage{fleqn,espcrc2}

\usepackage{graphicx}
\usepackage[figuresright]{rotating}

\newcommand{\AmS}{{\protect\the\textfont2
  A\kern-.1667em\lower.5ex\hbox{M}\kern-.125emS}}

\title{Heavy Baryons from Lattice NRQCD}

\author{Nilmani Mathur\address{TRIUMF, 4004 Wesbrook Mall,
Vancouver, British Columbia, Canada V6T 2A3\\},
Randy Lewis\address{Department of Physics, University of Regina, Regina, 
SK, Canada S43 0A2\\} and R.M. Woloshyn$^{a}$}
\begin{document}

\begin{abstract}
The mass spectrum of heavy quark baryons has been computed on anisotropic lattices using quenched lattice nonrelativistic
    QCD. The mass splittings between spin-1/2 and spin-3/2 baryons are also calculated. Results are compared to those obtained
    by using a Dirac-Wilson action of the D234 type. Color hyperfine effects in heavy baryons are also discussed.

\end{abstract}

\maketitle

\section{INTRODUCTION}
Extraction of correct mass splittings between vector and pseudoscalar mesons
 remains a challenging problem in lattice QCD.
While quenched lattice QCD has been unable to extract correct mass splittings \cite{quen_res}, so far this problem persists even with full lattice QCD \cite{unquen_res}. One may thus ask whether lattice results for baryon mass splittings exhibit similar suppression compared to experimental numbers. So far, only a few lattice QCD results have been reported for heavy baryons \cite{hv1,hv2,hv3,hv4,hv5}, and there are only two works \cite{hv1,hv2} where heavy quarks are treated nonrelativistically.  In this work we report heavy baryon results by using a nonrelativistic heavy quark action and an improved light quark action on anisotropic lattices.

\vspace*{-0.10in}

\section{NUMERICAL SIMULATION}

\subsection{Actions}
For light quarks we use a Dirac-Wilson action of the D234 type \cite{alford}, which has been used previously and detailed in Refs. \cite{hv3,hv4}. Its leading classical errors are {\it cubic} in lattice spacing. The gauge action as well as the heavy quark NRQCD action is detailed in Ref. \cite{spwave}. The gauge action is tadpole improved and the leading classical error is {\it quartic} in lattice spacing. The Hamiltonian corresponding to the NRQCD action
is complete to $\mathcal{O}(1/M^3)$ in the classical continuum limit. 

\subsection{Simulation Details}
This work is done with two sets of quenched gauge configurations on anisotropic lattices (lattice sizes $12^{3}\times 32$ and $14^{3}\times 38\,$ at $\beta=2.1$ and $2.3$ respectively) with a bare aspect ratio $a_{s}/a_{t} = 2$. For $\beta=2.1$, we use 720 configurations and for $\beta = 2.3$, the number of configurations is 442. Two sets of bare masses are used for heavy quarks while four sets of hopping parameters are used for light quarks and they are listed in Table 1. Interpolating operators used here are 
same as in Ref. \cite{hv4,hv5},
and correlation functions are calculated using both local and smeared (gauge invariant 
smearing at the sink \cite{spwave,hv4,hv5}) propagators. The time window to extract mass is chosen in a way such that the ending time is large and the fit is stable with variation of both starting and ending time by a few time steps. The fit corresponding to lowest $\chi^{2}$ is considered in this window-span. The temporal lattice spacing and correspondingly the scale is fixed by extracting the $\rho$-meson mass and the strange mass is fixed from the $\phi$-meson mass \cite{hv4}. 
The charm mass is fixed by extracting the $\eta_c$ mass, whereas $B^{0}$ mass is used to fix the bottom mass. As in Ref. \cite{hv4}, light quark extrapolation is done by extrapolating the hadron masses as a function of 
$c_{0} + c_{2}m_\pi^{2} + c_{3}m_{\pi}^{3}$, where $m_{\pi}$ is the pion mass.

\begin{table}[h]
\caption{Summary of lattice parameters.}
\centerline{}
\begin{tabular}{l}
\hline
$\beta$ \quad\quad\,$\kappa$ (light quark)\quad\quad\quad\,\,\,\,\,{bare mass}\\
\hspace*{1.8in}\,\,$c$\quad\quad\quad\,\,\,\,\,$b$\\
\hline
\hspace*{-0.08in}2.1\,\,\,\,\,0.229,0.233,0.237,0.240\,\,\,\,\,1.2,1.5\quad\,\,\,\,5.0,6.0\\
\hspace*{-0.08in}2.3\,\,\,\,\,0.229,0.233,0.237,0.240\,\,\,\,\,1.04,1.24\,\,\,\,3.7,4.2\\
\hline
\end{tabular}
\end{table}

The kinetic mass of a state can be extracted by using the usual NRQCD relation \cite{spwave}
\begin{eqnarray}
M_{kin} &=& {2\pi^{2}\over {N_{s}^{2}\xi^{2}a_{t}[a_{t}(E_{p}-E_{0})]}},
\end{eqnarray}
which is derived from $E = {p^{2}\over 2M_{kin}}$ (see Ref. \cite{spwave} for details and clarification of notations). Mass differences  between two states ($H_{1}$ and $H_{2}$) with the same heavy quark can be obtained by taking the difference of their zero momentum simulation energy :
\begin{eqnarray}
M_{H_{1}} - M_{H_{2}}&=&E^{1}_{sim}(0) - E^{2}_{sim}(0).
\end{eqnarray}
It is always easier (smaller error) to extract heavy meson or baryon masses with heaviest light quark ($\kappa = 0.229$) rather than extracting their masses with lighter light quark ($\kappa = 0.233$ and higher). Moreover, we can calculate mass differences more accurately rather than masses, and so one can use mass differences to a extract mass.
For example,
one can calculate the meson mass by using the relation
\begin{eqnarray}
M(q_{l},Q) &=& M(q_h,H) - \Delta M
\end{eqnarray}
where 
\begin{eqnarray}
\Delta M = \Delta E = E(q_h,Q) - E(q_{l},Q).
\end{eqnarray}
Here $q_{l}$ and $q_{h}$ denote lighter and heaviest light quark respectively.
This equation is valid as long as the heavy quark $Q$ is same in both states.
Similarly, masses of singly and doubly heavy baryons can be extracted from meson masses by using
\begin{eqnarray}
M(q_{1}q_{2},Q)     &=& M(q_{h},Q) - \Delta E_{sh},\\
M(q_{1},QQ)     &=& M(QQ) - \Delta E_{dh},
\end{eqnarray}
where
\begin{eqnarray}
E_{sh}   &=& E(q_{h},Q) - E(q_{1}q_{2},Q),\\
E_{dh}    &=&  E(QQ) - E(q_{1},QQ).
\end{eqnarray}
For example, $\Sigma_{c(b)}$ mass is extracted by taking its difference (at each $\kappa$) with  $D(B^{0})$ mass ($m$) at $\kappa = 0.229$ and then subtracting that from $m$. Masses extracted by using Eq. 1 and Eqs. 3-8 are consistent with each other. However, errors in the second method are considerably smaller than the previous one.

\begin{table}[t]
\caption{NRQCD Results (in units of MeV). First error is statistical while second error comprises systematic errors due to scale, time window and anisotropy.}
\centerline{}
\begin{tabular}{ccc}
\hline
&$\beta = 2.1$&$\beta = 2.3$\\
\hline
$\Sigma_{c}$&$2407(32)(^{32}_{37})$&$2471(38)(^{59}_{55})$\\
$\Omega_{c}$&$2652(25)(^{27}_{31})$&$2690(33)(^{52}_{48})$\\
$\Sigma^{*}_{c} - \Sigma_{c}$ &$75(20)(^{14}_{12})$
&$90(18)(^{14}_{13})$\\
$\Xi^{*}_{c} - \Xi^{\prime}_{c}$ &$71(18)(^{12}_{9})$
&$86(16)(^{11}_{12})$\\
$\Omega^{*}_{c} - \Omega_{c}$&$65(13)(^{7}_{8})$
&$78(14)(^{9}_{10})$\\
$\Sigma_{c} - \Lambda_{c} $ 
&$128(28)(^{39}_{28})$& $155(38)(^{42}_{35})$\\
$\Xi_{c} - \Xi^{\prime}_{c}$ &$106(22)(^{26}_{19})$&\\
\hline
$\Xi_{cc}$&$3562(47)(^{27}_{25})$&$3580(67)(^{41}_{33})$\\
$\Omega_{cc}$&$3681(44)(^{17}_{19})$&$3692(62)(^{36}_{27})$\\
$\Xi^{*}_{cc} - \Xi_{cc}$ &$63(14)(^{9}_{7})$
&$72(11)(^{7}_{8})$\\
$\Omega^{*}_{cc} - \Omega_{cc}$&$56(8)(^{7}_{6})$
&$65(7)(^{7}_{4})$\\
\hline
$\Lambda_{b}$&&$5681(102)(^{48}_{50})$\\
$\Sigma^{*}_{b} - \Sigma_{b}$ &$22(12)(^{8}_{6})$&$26(13)(^{8}_{6})$\\
$\Xi^{*}_{b} - \Xi^{\prime}_{b}$ &$21(12)(^{8}_{6})$&$26(12)(^{8}_{5})$\\
$\Omega^{*}_{b} - \Omega_{b}$&$18(8)(^{6}_{4})$&$24(8)(^{6}_{3})$\\
$\Lambda_{b} - \Sigma_{b}$ &$142(28)(^{30}_{22})$&\\
$\Xi_{b} - \Xi^{\prime}_{b}$ &$122(22)(^{32}_{18})$&\\
\hline
$\Xi^{*}_{bb} - \Xi_{bb}$ &$22(6)(^{4}_{3})$
&$21(6)(^{3}_{4})$\\
$\Omega^{*}_{bb} - \Omega_{bb}$&$20(4)(^{3}_{2})$
&$20(4)(^{3}_{3})$\\
$\Xi^{\prime}_{cb}$ & $6810(150)(^{62}_{79})$
& $6817(230)(^{69}_{121})$\\
$\Omega^{\prime}_{cb}$& $6935(135)(^{75}_{89})$
& $6928(217)(^{62}_{115})$\\
$\Xi^{*}_{cb} - \Xi^{\prime}_{cb}$ &$46(8)(^{4}_{6})$
&$48(9)(^{7}_{8})$\\
$\Omega^{*}_{cb} - \Omega^{\prime}_{cb}$&$40(6)(^{4}_{5})$
&$43(6)(^{6}_{6})$\\
$\Xi_{cb} - \Xi^{\prime}_{cb}$ &$11(6)(^{4}_{5})$&$10(5)(^{6}_{4})$\\
$\Omega_{cb} - \Omega^{\prime}_{cb}$&$10(5)(^{4}_{4})$&$10(4)(^{5}_{4})$\\
\hline
\end{tabular}
\end{table}

\begin{table}[t]
\caption{Comparison of NRQCD results with D234 results \cite{hv4} and experiment \cite{expt}. For lattice results, first row is for D234 while the second row is for NRQCD results (all are in MeV).}
\centerline{}
\begin{tabular}{cccc}
\hline 
&\multicolumn{2}{c}{Lattice results}&Expt\\
\hline
&$\beta =2.1$&$\beta =2.3$&\\
\hline
$\Sigma _{c}$&$2379(31)(_{18}^{23})$  
&$2490(14)(_{33}^{17})$&2455\\
&$2407(32)(^{32}_{37})$&$2471(38)(^{59}_{55})$&\\
$\Omega _{c}$
&$2671(11)(_{59}^{11})$
&$2699(10)(_{41}^{\, 8})$&2704\\
&$2652(25)(^{27}_{31})$&$2690(33)(^{52}_{48})$&\\
$\Sigma _{c}^{*}-\Sigma _{c}$
&$62(33)(_{32}^{19})$
&$82(12)(_{6}^{9})$&64\\
&$75(20)(^{14}_{12})$&$90(18)(^{14}_{13})$&\\
$\Xi _{c}^{*}-\Xi _{c}'$
&$52(15)(_{4}^{8})$
&$82(10)(_{5}^{8})$&70\\
&$71(18)(^{12}_{9})$&$86(16)(^{11}_{12})$&\\
$\Omega _{c}^{*}-\Omega _{c}$
&$50(17)(_{\, 6}^{11})$
&$73(8)(_{5}^{7})$&\\
&$65(13)(^{7}_{8})$&$78(14)(^{9}_{10})$&\\
$\Xi _{cc}$
&$3608(15)(_{35}^{13})$ 
&$3595(12)(_{22}^{21})$&\\
&$3562(47)(^{27}_{25})$&$3580(67)(^{41}_{33})$&\\
$\Omega _{cc}$
&$3747(9)(_{47}^{11})$
&$3727(9)(_{40}^{16})$&\\
&$3681(44)(^{17}_{19})$&$3692(62)(^{36}_{27})$&\\
$\Xi _{cc}^{*}-\Xi _{cc}$
&$58(14)(_{10}^{16})$
&$83(8)(_{10}^{\, 7})$&\\
&$63(14)(^{9}_{7})$&$72(11)(^{7}_{8})$&\\
$\Omega _{cc}^{*}-\Omega _{cc}$
&$57(8)(_{\, 9}^{10})$
&$72(5)(_{5}^{4})$&\\
&$56(8)(^{7}_{6})$&$65(7)(^{7}_{4})$&\\
\hline 
\end{tabular}
\end{table}

\section{RESULTS AND SUMMARY}
The mass spectrum and spin splittings of heavy quark baryons have been computed on anisotropic lattices using an NRQCD heavy quark action. 
Results are summarized in Table 2, whereas in Table 3 we have compared our results with those obtained by using a relativistic heavy quark (D234 action, \cite{hv4}) and experimental numbers (where available).
One can notice  NRQCD results and D234 results are consistent with each other.
Results are also consistent with a previous NRQCD calculation \cite{hv2}.

As it has been argued earlier \cite{hv4,hv5}, suppression of spin splittings is {\it not present} in baryon sector in the same way as it is in heavy meson sector (characteristic of quenched QCD simulations). 
\begin{figure}[t]
\vspace*{1.4in}
\includegraphics{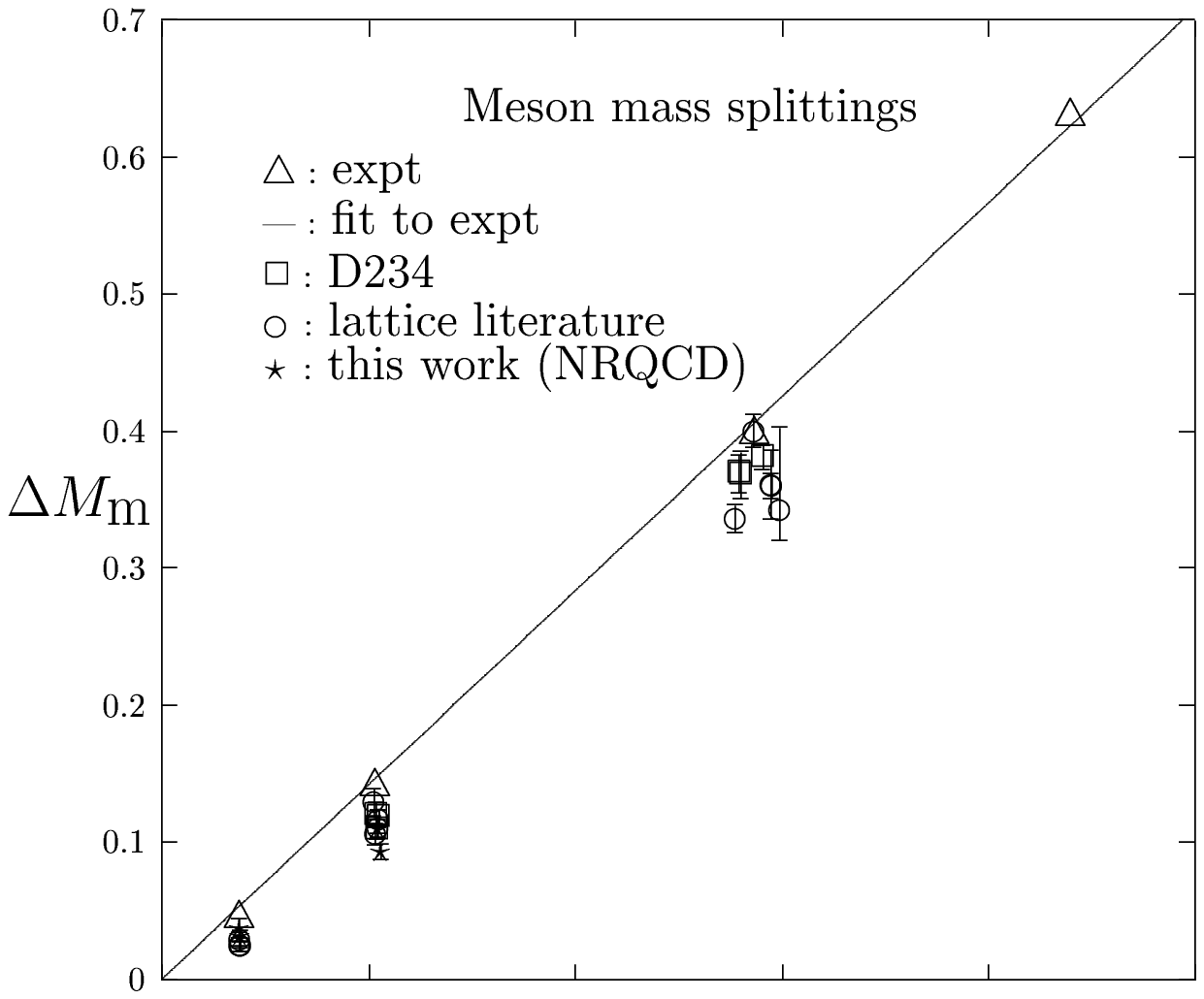}
\vskip -0.4in
\end{figure}
\begin{figure}
\includegraphics{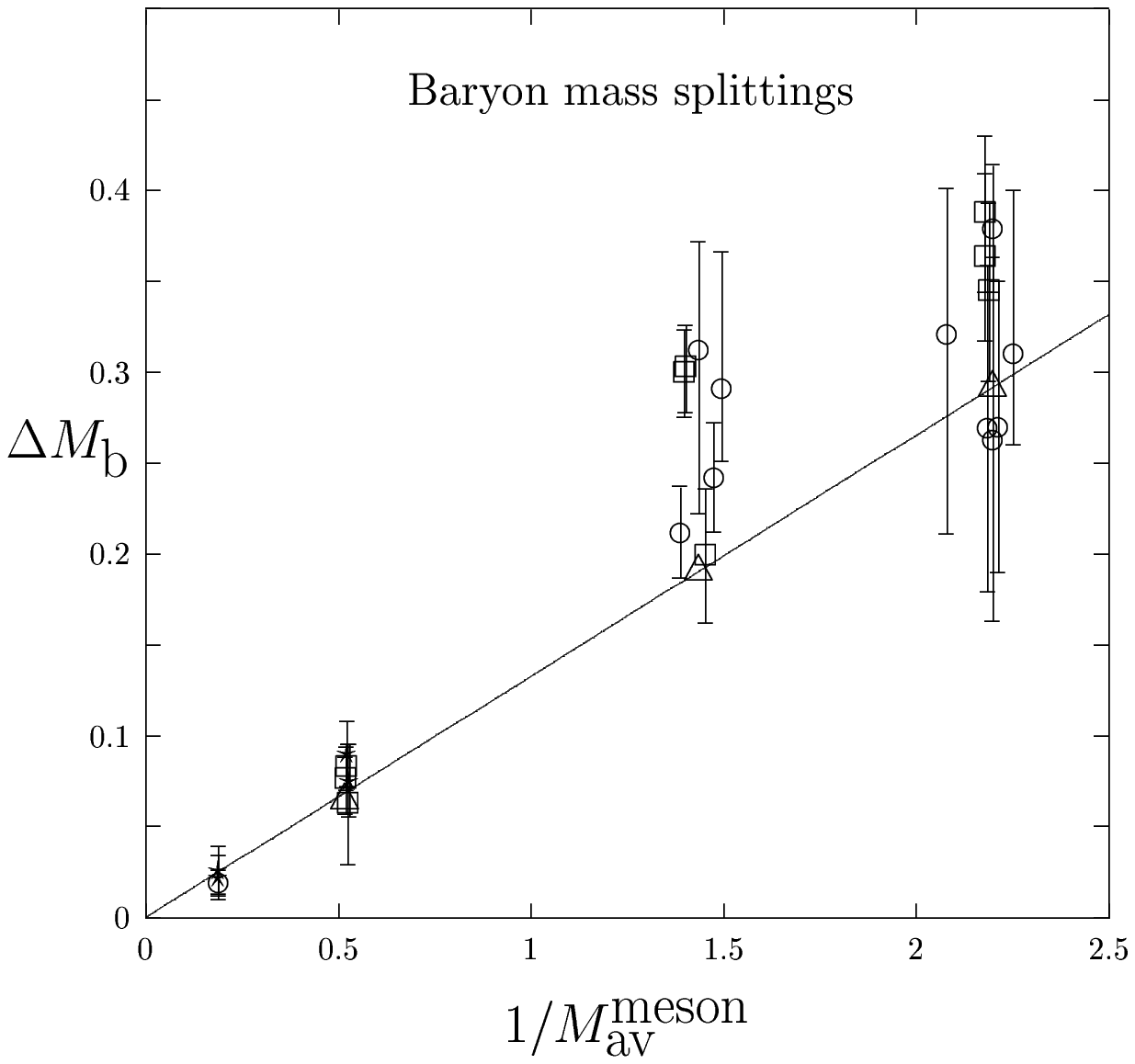}
\vskip 2.3in
\caption{Meson (top) and baryon (bottom) mass splittings as a function of average meson mass.}
\end{figure}
In Fig. 1 , we plot meson and baryon spin splittings respectively. There is a clear suppression in meson sector (top fig) which is not present in baryon sector (bottom fig).
Experimental measurement of spin splittings in bottom baryons as well as for doubly heavy baryons will be very useful to check if this is true indeed.
\rule{74mm}{0.2mm}

This work is supported by the Natural Sciences and Engineering Research Council of Canada.

\end{document}